\documentclass[twoside,twocolumn,9pt]{article}
\usepackage{extsizes}
\usepackage[super,sort&compress,comma]{natbib} 
\usepackage[version=3]{mhchem}
\usepackage[left=1.5cm, right=1.5cm, top=1.785cm, bottom=2.0cm]{geometry}
\usepackage{balance}
\usepackage{times,mathptmx}
\usepackage{sectsty}
\usepackage{graphicx} 
\usepackage{lastpage}
\usepackage[format=plain,justification=raggedright,singlelinecheck=false,font={stretch=1.125,small,sf},labelfont=bf,labelsep=space]{caption}
\usepackage{float}
\usepackage{fancyhdr}
\usepackage{fnpos}
\usepackage[english]{babel}
\usepackage{array}
\usepackage{droidsans}
\usepackage{charter}
\usepackage[T1]{fontenc}
\usepackage[usenames,dvipsnames]{xcolor}
\usepackage{setspace}
\usepackage[compact]{titlesec}
\usepackage{multirow}

\usepackage{epstopdf}%This line makes .eps figures into .pdf - please comment out if not required.

\definecolor{cream}{RGB}{222,217,201}

\begin{document}

\pagestyle{fancy}
\thispagestyle{plain}
\fancypagestyle{plain}{

%%%HEADER%%%
\fancyhead[C]{\includegraphics[width=18.5cm]{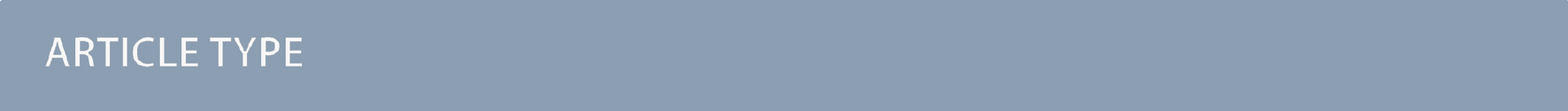}}
\fancyhead[L]{\hspace{0cm}\vspace{1.5cm}\includegraphics[height=30pt]{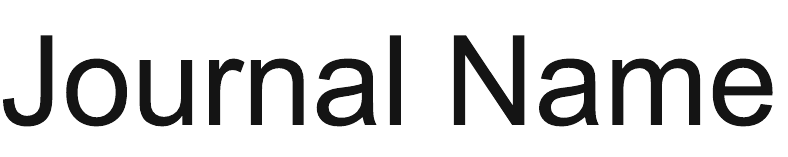}}
\fancyhead[R]{\hspace{0cm}\vspace{1.7cm}\includegraphics[height=55pt]{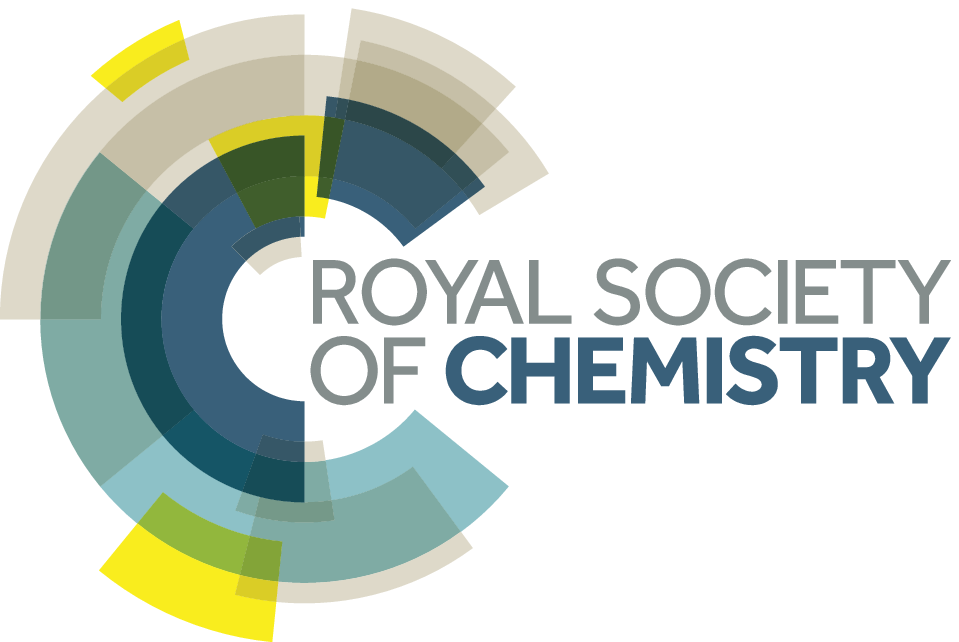}}
\renewcommand{\headrulewidth}{0pt}
}
%%%END OF HEADER%%%

%%%PAGE SETUP - Please do not change any commands within this section%%%
\makeFNbottom
\makeatletter
\renewcommand\LARGE{\@setfontsize\LARGE{15pt}{17}}
\renewcommand\Large{\@setfontsize\Large{12pt}{14}}
\renewcommand\large{\@setfontsize\large{10pt}{12}}
\renewcommand\footnotesize{\@setfontsize\footnotesize{7pt}{10}}
\makeatother

\renewcommand\floatpagefraction{.99}
\renewcommand\topfraction{.99}
\renewcommand\bottomfraction{.99}
\renewcommand\textfraction{.01}
\renewcommand\dbltopfraction{0.99}
\renewcommand\dblfloatpagefraction{0.99}

\renewcommand{\thefootnote}{\fnsymbol{footnote}}
\renewcommand\footnoterule{\vspace*{1pt}% 
\color{cream}\hrule width 3.5in height 0.4pt \color{black}\vspace*{5pt}} 
\setcounter{secnumdepth}{5}

\makeatletter 
\renewcommand\@biblabel[1]{#1}            
\renewcommand\@makefntext[1]% 
{\noindent\makebox[0pt][r]{\@thefnmark\,}#1}
\makeatother 
\renewcommand{\figurename}{\small{Fig.}~}
\sectionfont{\sffamily\Large}
\subsectionfont{\normalsize}
\subsubsectionfont{\bf}
\setstretch{1.125} %In particular, please do not alter this line.
\setlength{\skip\footins}{0.8cm}
\setlength{\footnotesep}{0.25cm}
\setlength{\jot}{10pt}
\titlespacing*{\section}{0pt}{4pt}{4pt}
\titlespacing*{\subsection}{0pt}{15pt}{1pt}
%%%END OF PAGE SETUP%%%

%%%FOOTER%%%
\fancyfoot{}
\fancyfoot[LO,RE]{\vspace{-7.1pt}\includegraphics[height=9pt]{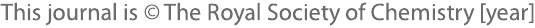}}
\fancyfoot[CO]{\vspace{-7.1pt}\hspace{13.2cm}\includegraphics{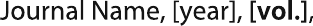}}
\fancyfoot[CE]{\vspace{-7.2pt}\hspace{-14.2cm}\includegraphics{head_foot/RF}}
\fancyfoot[RO]{\footnotesize{\sffamily{1--\pageref{LastPage} ~\textbar  \hspace{2pt}\thepage}}}
\fancyfoot[LE]{\footnotesize{\sffamily{\thepage~\textbar\hspace{3.45cm} 1--\pageref{LastPage}}}}
\fancyhead{}
\renewcommand{\headrulewidth}{0pt} 
\renewcommand{\footrulewidth}{0pt}
\setlength{\arrayrulewidth}{1pt}
\setlength{\columnsep}{6.5mm}
\setlength\bibsep{1pt}
%%%END OF FOOTER%%%

%%%FIGURE SETUP - please do not change any commands within this section%%%
\makeatletter 
\newlength{\figrulesep} 
\setlength{\figrulesep}{0.5\textfloatsep} 

\newcommand{\topfigrule}{\vspace*{-1pt}% 
\noindent{\color{cream}\rule[-\figrulesep]{\columnwidth}{1.5pt}} }

\newcommand{\botfigrule}{\vspace*{-2pt}% 
\noindent{\color{cream}\rule[\figrulesep]{\columnwidth}{1.5pt}} }

\newcommand{\dblfigrule}{\vspace*{-1pt}% 
\noindent{\color{cream}\rule[-\figrulesep]{\textwidth}{1.5pt}} }

\makeatother
%%%END OF FIGURE SETUP%%%

%%%TITLE, AUTHORS AND ABSTRACT%%%
\twocolumn[
  \begin{@twocolumnfalse}
\vspace{3cm}
\sffamily
\begin{tabular}{m{4.5cm} p{13.5cm} }

\includegraphics{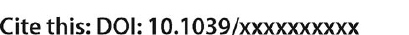} & \noindent\LARGE{\textbf{Large elastic recovery of zinc dicyanoaurate$^\dag$}} \\%Article title goes here instead of the text "This is the title"
\vspace{0.3cm} & \vspace{0.3cm} \\

 & \noindent\large{Chloe S. Coates,\textit{$^{a}$} Matthew R. Ryder,\textit{$^{b}$} Joshua A. Hill,\textit{$^{a}$} Jin-Chong Tan,$^\ast$\textit{$^{b}$} and}\\
 &\noindent\large{Andrew~L.~Goodwin$^{\ast}$\textit{$^{a}$}} \\%Author names go here instead of "Full name", etc.

\includegraphics{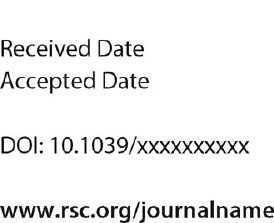} & \noindent\normalsize{We report the mechanical properties of the `giant' negative compressibility material zinc(II) dicyanoaurate, as determined using a combination of single-crystal nanoindentation measurements and \emph{ab initio} density functional theory calculations. While the elastic response of zinc dicyanoaurate is found to be intermediate to the behaviour of dense and open framework structures, we discover the material to exhibit a particularly strong elastic recovery, which is advantageous for a range of practical applications. We attribute this response to the existence of supramolecular helices that function as atomic-scale springs, storing mechanical energy during compressive stress and hence inhibiting plastic deformation. Our results are consistent with the relationship noted in [Cheng \& Cheng, \textit{Appl. Phys. Lett.}, 1998, {\textbf{73}}, 614] between the magnitude of elastic recovery, on the one hand, and the ratio of material hardness to Young's modulus, on the other hand. Drawing on comparisons with other metal--organic frameworks containing helical structure motifs, we suggest helices as an attractive supramolecular motif for imparting resistance to plastic deformation in the design of functional materials.}
\end{tabular}

 \end{@twocolumnfalse} \vspace{0.6cm}

  ]
%%%END OF TITLE, AUTHORS AND ABSTRACT%%%

%%%FONT SETUP - please do not change any commands within this section
\renewcommand*\rmdefault{bch}\normalfont\upshape
\rmfamily
\section*{}
\vspace{-1cm}

%%%FOOTNOTES%%%

\footnotetext{\textit{$^{a}$~Department of Chemistry, University of Oxford, Inorganic Chemistry Laboratory, South Parks Road, Oxford OX1 3QR, U.K. Tel: +44 (0)1865 272137; E-mail: andrew.goodwin@chem.ox.ac.uk}}
\footnotetext{\textit{$^{b}$~Department of Engineering Science, University of Oxford, Parks Road, Oxford OX1 3PJ, U.K. Tel: +44 (0)1865 273925; E-mail: jin-chong.tan@eng.ox.ac.uk}}

%Please use \dag to cite the ESI in the main text of the article.
%If you article does not have ESI please remove the the \dag symbol from the title and the footnotetext below.
\footnotetext{\dag~Electronic Supplementary Information (ESI) available: details of \emph{ab initio} calculations; nanoindentation load--displacement data and detailed analysis; optical microscopy. See DOI: 10.1039/b000000x/}
%additional addresses can be cited as above using the lower-case letters, c, d, e... If all authors are from the same address, no letter is required

%\footnotetext{\ddag~Additional footnotes to the title and authors can be included \emph{e.g.}\ `Present address:' or `These authors contributed equally to this work' as above using the symbols: \ddag, \textsection, and \P. Please place the appropriate symbol next to the author's name and include a \texttt{\textbackslash footnotetext} entry in the the correct place in the list.}

%%%END OF FOOTNOTES%%%

%%%MAIN TEXT%%%%

\section{Introduction}
%Hybrid inorganic-organic framework materials exhibit a range of extraordinary mechanical properties, often as a consequence of their extreme atomic-scale flexibility. 

Framework materials---solids with structures assembled from a combination of nodes and linkers\cite{Hoskins1990,Eddaoudi_2001}---are well-known to harbour a variety of interesting mechanical properties. This is especially true for systems with long and flexible linkers since flexibility generally amplifies mechanical response.\cite{Coudert_2015} One topical example is the broad family of metal--organic frameworks (MOFs), members of which are known to exhibit \emph{e.g.}\ anomalously low elastic moduli,\cite{Tan2012,Ryder_2016} extreme mechanical anisotropy,\cite{Ortiz2012} negative Poisson's ratios,\cite{Greaves2011} and a propensity for negative linear compressibility (NLC).\cite{Baughman1998,Li2014,Cairns_2015} Taken at face value, these properties suggest a range of attractive applications for open frameworks such as MOFs and MOF-like systems, including in stimuli-responsive materials (sensors), actuators, and shock absorbers.\cite{Coudert_2015,Aliev2009,Spinks2002}

Fundamental studies of framework mechanical response often (rightly) focus on elastic behaviour, and yet the relevance of the elastic regime to practical applications relies on the extent to which plastic deformation can be avoided under application of stress. Often the origin of this stress is an external pressure (\emph{e.g.}\ hydrostatic or uniaxial), but guest sorption is an alternative mechanism of particular relevance to MOFs.\cite{Krause2016} An important measure of resistance to irreversible deformation is the elastic recovery of a material, $W_{\textrm e}$, which quantifies the proportion of work resulting in elastic deformation.\cite{Cheng1998} Put simply, materials with high elastic recoveries will regain their structure and shape after stress cycling: the elastic recovery of rubber, for example, is nearly unity, whereas brittle ceramics have values of $W_{\textrm e}$ close to zero. Consequently there is clear merit in developing an understanding of the elastic recovery of open frameworks so as to help predict the extent to which their anomalous mechanical responses might be exploited in practice.

Here we study the elastic recovery of zinc(II) dicyanoaurate(I), Zn[Au(CN)$_2$]$_2$, a MOF-like system known for its extreme NLC behaviour.\cite{Cairns2013} Whereas the uniaxial compressibilities $K_{\ell}=-(\partial\ell/\ell\,\partial p)_T$ of conventional materials are usually in the range $\sim5$--$10$\,TPa$^{-1}$ (\emph{i.e.}\ a linear contraction of 0.5--1\% for each 1\,GPa of applied pressure),\cite{Zhang2010} Zn[Au(CN)$_2$]$_2$ exhibits a remarkably large and negative compressibility along the hexagonal axis of its quartzlike structure, with $K_c=-42(5)$\,TPa$^{-1}$. In other words, its structure \emph{expands} on compression by roughly 4\% for each GPa. This extraordinary elastic property places Zn[Au(CN)$_2$]$_2$ in a unique position for application in pressure sensors and shock amplifiers.\cite{Nicolaou2012} Yet remarkably little is known regarding its mechanical behaviour beyond its (elastic) compressibility and thermal expansivity.\cite{Cairns2013,Goodwin2009} As an experimental technique, nanoindentation methods have provided crucially important insight into the mechanical properties of flexible MOFs.\cite{Best2015, Qu2016, Hobday2016} Consequently our approach is to use a combination of nanoindentation measurements and \emph{ab initio} density functional theory (DFT) calculations to determine the experimental Young's moduli, hardness, and elastic recovery of Zn[Au(CN)$_2$]$_2$ and then place these values in the context of its full elastic tensor, as determined computationally. The most important result of our study is the discovery of a large elastic recovery, which we attribute to the same supramolecular motifs thought to be responsible for NLC itself.

\begin{figure}
\centering
\includegraphics{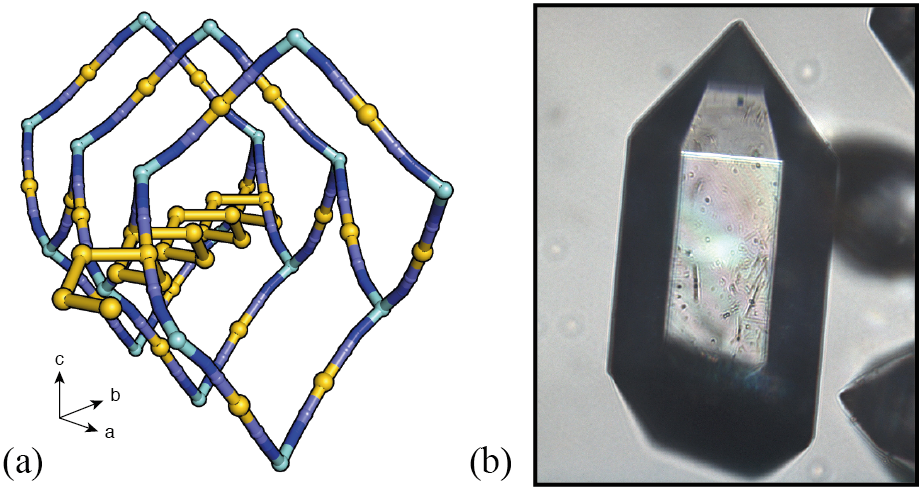}% Here is how to import EPS art
\caption{\label{fig1} (a) Representation of a fragment of the crystal structure of Zn[Au(CN)$_2$]$_2$. Zn atoms in green, C and N atoms in blue, and Au atoms in gold. Only one of the six interpenetrating quartzlike frameworks is shown. Aurophilic interactions between neighbouring frameworks form helices oriented perpendicular to the hexagonal $c$-axis, which is the vertical direction in this representation. (b) The compound crystallises as elongated hexagonal bipyramids, occasionally truncated. The well-defined crystal facets make Zn[Au(CN)$_2$]$_2$ an ideal candidate for nanoindentation studies.}
\end{figure}

We conclude this introduction with a brief summary of the structural chemistry of Zn[Au(CN)$_2$]$_2$. The crystal structure was first reported in Ref.~\citenum{Hoskins1995}, where the relationship to the structure of SiO$_2$-quartz was clearly noted [Fig.~\ref{fig1}(a)]. Like that of $\beta$-quartz, the crystal symmetry of Zn[Au(CN)$_2$]$_2$ is hexagonal, and so from a mechanical perspective the important crystal directions are the [100] and [001] axes. The quartzlike nets of Zn[Au(CN)$_2$]$_2$ are assembled from tetrahedral Zn$^{2+}$ centres, connected \emph{via} almost-linear dicyanoaurate ([Au(CN)$_2$]$^-$) linkers. In this structure, the Zn$\ldots$Zn distance is so large (\emph{ca} 1\,nm) that six separate quartzlike nets interpenetrate.\cite{Hoskins1995,Batten2001,Hill2016} While each net is covalently distinct, neighbouring nets are thought to interact via aurophilic (Au$\ldots$Au) interactions,\cite{Hoskins1995} the existence of which is inferred from the relatively short Au$\ldots$Au separations observed crystallographically.\cite{Jansen_1987,Pyykko_1997,Schmidbaur_2000,Katz2008} These interactions connect to form a set of aurophilic helices lying along directions perpendicular to the hexagonal axis. Variable-pressure crystallographic measurements suggest that the extreme compressibility of this system arises because these aurophilic helices function as supramolecular `springs'.\cite{Baughman1998} Just as a steel spring is more compressible than steel itself, so too are the aurophilic helices in Zn[Au(CN)$_2$]$_2$ more compressible than the Au$\ldots$Au interactions from which they are made. We will come to show that this same supramolecular motif is likely responsible for the strong elastic recovery of the material.

The remainder of our paper is arranged as follows. In section \ref{methods} we summarise the experimental and computational methods used in our study. We then present the results of our nanoindentation experiments, and compare these with the elastic response anticipated from our DFT calculations. In this process, we include a discussion regarding the determination of elastic recovery using nanoindentation methods. Finally, we place the elastic recovery of Zn[Au(CN)$_2$]$_2$ in the context of the behaviour of other dense and open frameworks, highlighting the existence and nature of key supramolecular motifs that favour high recovery values.

%In this paper we report the elastic modulus, hardness and elastic recovery for the (100) and (001) facets in zinc dicyanoaurate. We find, in addition to significant anisotropy in the elastic constants, that the elastic recovery is particularly strong along [100]. We suggest that this large elastic recovery is related to the existence of the supramolecular spring motif and we present empirical evidence for related structural features and elastic recovery in framework materials.

\section{Methods}\label{methods}

\subsection*{Synthesis and characterisation}
Single-crystal samples of zinc(II) dicyanoaurate(I) were prepared using the same hydrothermal synthesis approach reported in Ref.~\citenum{Cairns2013}. In a 23\,mL Teflon-lined Parr reaction vessel, we combined aqueous solutions of Zn(CH$_3$COO)$_2\cdot2$H$_2$O (Sigma Aldrich, 98\%, 14\,mg) and KAu(CN)$_2$ (Sigma Aldrich, 98\%, 60\,mg) such that the total volume of liquid was approximately 10\,mL. The vessel was sealed, heated to 120\,$^\circ$C for 2\,h, and then cooled at a rate of 1\,$^\circ$C\,h$^{-1}$ to ambient temperature. On inspection, the reaction mixture contained large single crystals of Zn[Au(CN)$_2$]$_2$ in the form of elongated hexagonal bipyramids [Fig.~\ref{fig1}(b)]. This is the same morphology reported in Refs.~\citenum{Goodwin2009,Cairns2013} and corresponds to the $\alpha$-polymorph as described in Ref.~\citenum{Katz2008}. A suitable crystal was mounted on an Oxford Diffraction (Rigaku Oxford Diffraction) SuperNova diffractometer, and a sufficient portion of the X-ray diffraction pattern was collected to enable face indexing using the Agilent Pro Crysalis Software.\cite{CrysalisPRO}The crystal was mounted using Paratone oil on a MiTeGen loop at ambient temperature.

\subsection*{Nanoindentation}
Nanoindentation studies were carried out at ambient temperature using an MTS NanoIndenter$^\copyright$ XP, equipped with a continuous stiffness measurement (CSM) module. The instrument was placed within an isolation cabinet that shielded against thermal instability and acoustic interference. A three-sided pyramidal Berkovich indenter with a sharp tip (end radius $\leq100$\,nm) was used to measure the indentation modulus and hardness.\cite{Berkovich1951} Calibration was performed using a fused silica standard, with elastic modulus ($E = 72$\,GPa) and hardness ($H = 9$\,GPa). Thermal drifts were ensured to be consistently low (below 0.1\,nm\,s$^{-1}$).  Indentation depth was 2000\,nm for all measurements and the inter-indent spacing was typically 35--40\,$\mu$m to avoid interference between neighbouring indentations. We probed two orientations of Zn[Au(CN)$_2$]$_2$ crystals, cleaved and polished so as to expose (100) and (001) crystal faces. Continuous stiffness measurement allowed us to collect depth-dependent mechanical data. The indentation modulus was determined using the Oliver and Pharr method;\cite{Oliver2004} the hardness was determined by the ratio 
\begin{equation}
H=\frac{p_{\textrm{max}}}{A_{\textrm{max}}},
\end{equation}
where $p_{\textrm{max}}$ and $A_{\textrm{max}}$ are, respectively, the applied force and contact area at maximum indentation.

%A useful experimental tool to quantify this elastic recovery is nanoindentation. Nanoindentation is a technique employed by materials scientists to obtain the Young's modulus and hardness of a material to characterise the  mechanical response of a framework to uniaxial pressure. We are able to extract the elastic recovery from the area under the load-displacement graphs as detailed below. The results of indentation experiments can be elegantly correlated to underlying chemistry and topology for a system. Extreme mechanical responses, such as the NLC observed in zinc dicyanoaurate, are typically a result of anisotropy in the crystal structure and so the mechanical properties extracted from nanoindentation experiments can give important information about the mechanism.

\subsection*{\emph{Ab initio} calculations}
%\emph{Ab initio} calculations were carried out using DFT, as implemented in the CRYSTAL14 software suite\cite{Dovesi2014}. Gaussian-type atom-centred all-electron basis sets were used for all atoms (\emph{were they?)}, giving a total of XXX basis functions. The atom coordinates and cell parameters of Zn[Au(CN)$_2$]$_2$ as reported in Ref.~\citenum{Cairns2013} were used as input and the crystal geometry was optimised at the B3LYP level. We considered relaxation to be complete once the root mean squared values of the gradient and displacement had converged to less than 0.0002\,a.u.\ and 0.0004\,a.u., respectively. The self-consistent field convergence threshold on the total energy was 10--8\,a.u. The elastic tensor constants $C_{ij}$ were then computed using the numerical first derivative of the analytic cell gradients\cite{ref}.
DFT calculations were carried out at the PBE level of theory,\cite{Perdew_1996} with 1428 Gaussian-type basis functions using the periodic \emph{ab initio} code CRYSTAL14.\cite{Dovesi_2014} Geometry optimisation at constant symmetry, with the full relaxation of both lattice parameters and atomic coordinates, was achieved \emph{via} a quasi-Newtonian algorithm. The experimental atomic coordinates and cell parameters of Zn[Au(CN)$_2$]$_2$ were used as the starting geometry,\cite{Goodwin2009} and the optimisation was considered to be complete when the root mean squared values of the gradient and atomic displacement had converged simultaneously to less than 0.0001 and 0.0002 a.u., respectively. The elastic constants were obtained from the optimised structure by first calculating the single-point self-consistent-field (SCF) energy and then calculating the four sets of required strains (due to Zn[Au(CN)$_2$]$_2$ possessing hexagonal symmetry).  For each different directional strain, the structure was deformed, and the new symmetry elements were determined. This procedure was carried out for multiple strain steps to increase the accuracy of the resultant gradient. For each deformed structure, the atomic coordinates were relaxed and optimised as above. A further SCF energy calculation was then performed at each optimised deformation,  the energy gradient was fitted with singular-value-decomposition routines, and the second derivatives were determined numerically.\cite{Perger_2009} This approach allowed the elastic constants to be computed and the mechanical properties to be obtained \emph{via} tensorial analysis.\cite{Marmier_2010b}

\section{Results and Discussion}
\begin{figure}
\includegraphics{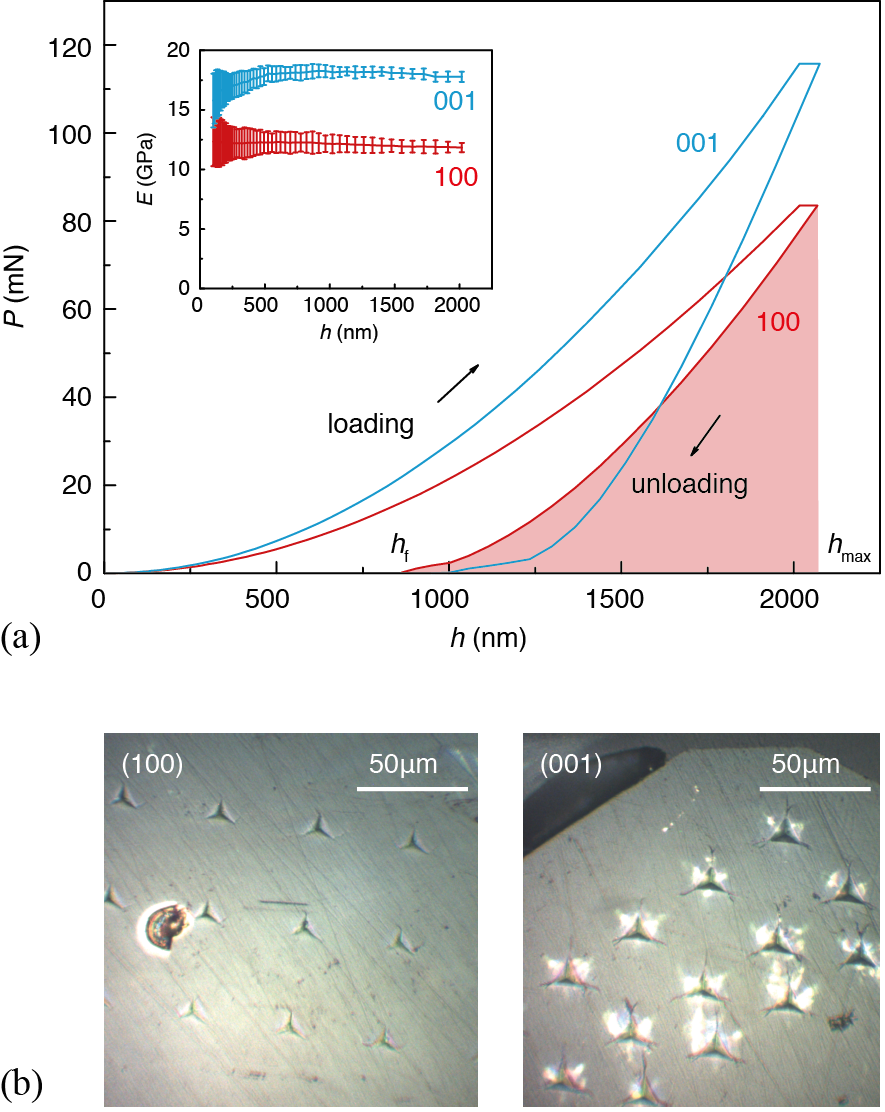}% Here is how to import EPS art
\caption{\label{fig2} Typical load--displacement plots obtained from nanoindentation measurements of Zn[Au(CN)$_2$]$_2$ using the (100) and (001) facets. The elastic recovery $W_{\textrm e}$ can be calculated from the ratio of the areas under the unloading and loading curves, as described in the text. The inset shows the effective elastic moduli extracted from these data, given as a function of indentation depth $h$. (b) Microscopy images of the (100) and (001) surfaces (left and right, respectively) following nanoindentation to a common maximum penetration depth $h_{\textrm{max}}$. Note that the degree of residual deformation is much reduced for (100) than for (001); this observation is consistent with the greater elastic recovery of the former relative to the latter.}
\end{figure}

Typical load--displacement curves for Zn[Au(CN)$_2$]$_2$ as obtained \emph{via} nanoindentation are shown in Figure~\ref{fig2}(a). The particular shape of these curves reflects the pyramidal nature of the Berkovich tip used in our study, which results in an increasing contact area at higher loads---and hence decreasing variation in displacement depth on indentation. The behaviour we observe is characteristic of the nanoindentation response of framework materials, showing contributions from both elastic and plastic deformation;\cite{Oliver2004} the latter is evident in that loading and unloading curves are not coincident. We begin our analysis by focusing on the elastic response of Zn[Au(CN)$_2$]$_2$: we extract elastic moduli from our nanoindentation data and compare these to the results of \emph{ab initio} calculations. We then proceed to report the hardness values and compare the relative fractions of reversible (elastic) and irreversible (plastic) work done during nanoindentation to establish the elastic recovery of Zn[Au(CN)$_2$]$_2$ in different crystal orientations.

Indentation moduli were determined using continuous stiffness measurements on loading, using data collected for at least seventeen indents for both (100) and (001) crystal faces [inset to Figure~\ref{fig2}(a); see ESI for further discussion]. Using a representative value of the effective isotropic Poisson's ratio ($\nu=0.2$), we obtain the values $E_{100}=14.5(3)$\,GPa and $E_{001}=21.4(5)$\,GPa; these values vary by not more than 30\% when recalculated with alternate values of $\nu$ (see ESI). %which are consistent with the bulk modulus determined using variable-pressure single-crystal x-ray diffraction measurements ($B_0=16.7$\,GPa)\cite{Cairns2013}.
%Indentation moduli are not direct measurements of the Young's moduli, but we treat the former as a useful approximation to the latter for the remainder of our analysis (\emph{e.g.}\ as in Ref.~\citenum{Oliver2004}).
The moduli are low in comparison with corresponding values for many other coordination polymers\cite{Zhang2016} but remain higher than the values reported for most porous MOFs.\cite{Ortiz2012,Coudert_2015,Bahr2007} Zn[Au(CN)$_2$]$_2$ is not porous, and yet its number density ($\rho=0.0512$\,atoms\,\AA$^{-3}$) shares more in common with open framework materials such as the zeolitic imidazolate framework (ZIF) family of MOFs\cite{Park_2006,Banerjee_2008} ($\rho<0.08$\,atoms\,\AA$^{-3}$) than it does with non-porous frameworks ($\rho>0.08$\,atoms\,\AA$^{-3}$). Hence, we suggest that the mechanical behaviour of Zn[Au(CN)$_2$]$_2$ is MOF-like, despite the absence of any appreciable pore volume in its structure. This borderline behaviour is consistent with the observation that the isostructural material Zn[Ag(CN)$_2$]$_2$ is known to act as a host for the inclusion of inorganic guests.\cite{Goodwin2009,Hill2016}

The elastic response of Zn[Au(CN)$_2$]$_2$ shows clear evidence of anisotropy: the elastic modulus along [100] is 32\% lower than along [001]. This anisotropy is consistent with the orientation within the crystal structure of aurophilic helices---supramolecular motifs that might be expected to be particularly compliant. 

%, and hardness values of $H_{100}=2.23(4)$\,GPa and $H_{001}=2.19(7)$\,GPa

% which corresponds to an anisotropy of 1:1.32.

%This elastic anisotropy is perhaps to be expected on the basis of the structure of Zn[Au(CN)$_2$]$_2$. It is comprised of six interpenetrated $\beta$-quartz nets which form honeycomb-like channels perpendicular to the $\mathbf{c}$-axis. The honeycomb topology is known for its ability to accommodate uniaxial compression via expansion in the perpendicular direction, hence why the [001] is relatively compliant. This perpendicular expansion is also responsible for the even lower value of $E_{100}$, but in this case it is coupled to the compression of the helix of gold atoms along the [100] which lowers the elastic constant still further. 

Our \emph{ab initio} calculations show good agreement with these experimental results. A summary of the relaxed cell geometry and 0\,K elastic properties determined computationally is given in Table~\ref{table1}. The small differences between calculated and measured moduli $E$ likely reflect the fact that nanoindentation measurements probe an effective modulus that is only an approximation to the Young's modulus.\cite{Oliver2004} Nevertheless the nature and degree of mechanical anisotropy is clearly consistent between the two approaches: the calculated value of $E_{100}$ is lower by 29\% than that of $E_{001}$. Our calculations do not take into account dispersion effects, which can be important for frameworks containing metallophilic interactions.\cite{Schmidbaur_2000,OGrady2004,Schmidbaur_2015} We found that the dispersion-corrected B3LYP-D method, which includes damped empirical $C_6r^{-6}$ terms,\cite{Civalleri_2008} resulted in severe overbinding of the Zn[Au(CN)$_2$]$_2$ in the $\mathbf a$ and $\mathbf b$ directions such that the discrepancy between calculated and experimental lattice parameters was unacceptably large ($>15$\%). It is possible that full treatment of many-body dispersion effects\cite{Tkatchenko_2012} would provide the most accurate computational description of Zn[Au(CN)$_2$]$_2$---as is the case for the conceptually-related system Ag$_3$[Co(CN)$_6$] (Ref.~\citenum{Liu2016})---but this level of complexity is beyond the scope of our present study.

\begin{table}
\small
  \caption{\ Comparison between calculated and experimental structural and elastic properties of Zn[Au(CN)$_2$]$_2$}
  \label{table1}
  \begin{tabular*}{\columnwidth}{@{\extracolsep{\fill}}lccc}
\hline
& DFT (0\,K)& Expt. (298\,K)&Ref.\\ \hline
$a$ (\AA)&8.4880&8.4624(3)&\citenum{Cairns2013}\\
$c$ (\AA)&20.7255&20.6220(7)&\citenum{Cairns2013}\\
$V$ (\AA$^3$)&1293.14&1278.93(14)&\citenum{Cairns2013}\\ \hline
$K_a$ (TPa$^{-1}$)&51.02&55(16)&\citenum{Cairns2013}\\
$K_c$ (TPa$^{-1}$)&$-$49.32&$-$48(14)&\citenum{Cairns2013}\\
$B$ (GPa)&18.97$^\ast$&16.7(16)&\citenum{Cairns2013}\\
$E_{100}$ (GPa)&8.23&14.5(3)&This work\\
$E_{001}$ (GPa)&11.60&21.4(5)&This work\\ \hline
%$H_{\textrm{100}}$ (GPa)&--&2.23(4)&This work\\
%$H_{\textrm{001}}$ (GPa)&--&2.19(7)&This work\\ \hline
$E_{\textrm{max}}$ (GPa)&25.33&--\\
$E_{\textrm{min}}$ (GPa)&8.23&--\\
$G_{\textrm{max}}$ (GPa)&11.03&--\\
$G_{\textrm{min}}$ (GPa)&2.91&--\\
$\nu_{\textrm{max}}$&1.10&--\\
$\nu_{\textrm{min}}$&0.02&--&\\ \hline
  \end{tabular*}
\footnotesize{ $^\ast$This value corresponds to the direct modulus. We also calculated the Voigt, Reuss, and Hill moduli, which were equal to 59.14, 18.97, and 39.06\,GPa, respectively.}
\end{table}

Calculation offers insight into mechanical properties beyond those we can measure directly using nanoindentation. For example, the Young's modulus is clearly much more anisotropic than measurements along the [100] and [001] axes alone would suggest: the framework is substantially stiffer along the $\sim\langle$101$\rangle$ directions, giving a total elastic anisotropy of 3.1. This direction corresponds to the orientation of the Zn--NC--Au--CN--Zn framework struts, and is the same direction along which compressibility also vanishes (see ESI for further discussion).\cite{Cairns2013} Anisotropy is even more evident in the shear modulus, the minimum value of which (2.91\,GPa) suggests the material is not elastically stable far into the GPa regime. This is consistent with the experimental observation of a displacive phase transition at a hydrostatic pressure of 1.8\,GPa.\cite{Cairns2013} We note that the Poisson's ratio $\nu_{\mathbf{e}_1\mathbf{e}_2}$ vanishes for directions $\mathbf e_1,\mathbf e_2\in\langle$100$\rangle$ with $\mathbf e_1\perp\mathbf e_2$. One interpretation of this behaviour is the ability of aurophilic helices to absorb strain along their axes (parallel to $\langle$100$\rangle$ directions) without transmitting this strain in an orthogonal in-plane direction.

Analysis of load--displacement curves measured during nanoindentation also yields effective hardness values $H$ in addition to the moduli $E$. We find $H_{100}=2.23(4)$\,GPa and $H_{001}=2.19(7)$\,GPa, suggesting that hardness is essentially isotropic for Zn[Au(CN)$_2$]$_2$. The magnitude of these values is again intermediate to those for dense frameworks, on the one hand, and ZIFs, on the other hand.\cite{Tan2011}

The elastic recovery, $W_{\textrm e}$, was determined from our nanoindentation measurements by exploiting the relationship between the area under the loading or unloading curve and the work done on indentation or release, respectively.\cite{Cheng1998} The total work done during indentation includes both elastic and plastic contributions, and is given by the integral
\begin{equation}
W_{\textrm{total}}=\int^{h_{\textrm{max}}}_0P_{\uparrow}\,{\rm d}h,
\end{equation}
where $h_{\textrm{max}}$ is the maximum tip displacement, and $P_{\uparrow}(h)$ represents the load applied during the loading cycle at displacement $h$. The elastic component is determined by the equivalent integral evaluated using the unloading curve:
\begin{equation}
W_{\textrm{elastic}}=\int^{h_{\textrm{max}}}_{h_{\textrm f}}P_{\downarrow}\,{\rm d}h.
\end{equation}
Here $h_{\textrm f}$ is the final indenter depth [Fig.~\ref{fig2}(a)] and $P_{\downarrow}(h)$ represents the load applied during the unloading cycle at displacement $h$. The elastic recovery then simply corresponds to the ratio of these two work terms:\cite{Cheng1998, Miura2003}
\begin{equation}\label{we}
W_{\textrm e}=\frac{W_{\textrm{elastic}}}{W_{\textrm{total}}}.
\end{equation}
An alternate definition\cite{Xu2006} uses the ratio of the indentation depth recovered on unloading ($h_{\textrm{max}}-h_{\textrm f}$) to the final indentation depth on loading $h_{\textrm{max}}$:
\begin{equation}
W_{\textrm e}\sim\frac{h_{\textrm{max}}-h_{\textrm f}}{h_{\textrm{max}}}.
\end{equation}
This approach effectively assumes that the functions $P_{\uparrow}(h)$ and $P_{\downarrow}(h)$ have the same form but the latter is compressed in $h$-space by a factor $(h_{\textrm{max}}-h_{\textrm f})/h_{\textrm{max}}$. We find reasonable agreement between the two approaches but favour Eq.~\eqref{we} because of its more rigorous basis.

It is immediately apparent from Figure~\ref{fig2}(a) that the elastic recovery for indentation in the [100] direction is greater than for the [001] direction. The numerical values we obtain using Eq.~\eqref{we} are $W_{\textrm e}$ = 62.7(3)\% and 43.6(6)\% for the (100) and (001) faces, respectively. These values are consistent with the recovery observed directly by inspection of the crystal post-indentation, as shown explicitly in Fig.~\ref{fig2}(b). The (100) face shows a much higher resistance to irreversible deformation: the indents are clearly more shallow and show less cracking compared to those for the (001) face. A high elastic recovery leads to improved fracture toughness---\emph{i.e.}\ resistance to cracking.\cite{Tan2009}

% We note here that cracking can artificially affect the values of the elastic modulus and hardness (usually by decreasing them), which assume that the system is behaving within the elastic regime, hence our results for the modulus and hardness were obtained from crack-free indents, as discussed above. 

For context, we computed elastic recovery values for a range of other MOFs and coordination polymers using published data for the ZIF family\cite{Tan2010} as well as some dense hybrid frameworks: copper phosphonoacetate (polymorphs 1 and 2, referred to here as CuPA1 and CuPA2),\cite{Tan2009} cerium oxalate formate (CeOx),\cite{Tan2009a} and zinc phosphate phosphonoacetate hydrate (ZnPAA).\cite{Kosa2010} The relevant values are shown in Table~\ref{table2}. What is immediately clear is that the $W_{\textrm e}$ value we measure for Zn[Au(CN)$_2$]$_2$ along [100] is amongst the largest for all these systems, and is comparable to that of fused silica, which is widely used as a high-recovery standard in nanoindentation studies.\cite{Martin_2002} The behaviour of Zn[Au(CN)$_2$]$_2$ along [001] for Zn{[}Au(CN)$_2${]}$_2$ is typical of a dense hybrid framework, such that the system as a whole behaves like a MOF along $\mathbf{a}$, but like a dense framework along $\mathbf{c}$.

\begin{table}
\small
  \caption{\ Elastic moduli, hardnesses, and elastic recoveries for some dense and open framework materials}
  \label{table2}
  \begin{tabular*}{\columnwidth}{@{\extracolsep{\fill}}cccccc}
\hline
{Compound }              &    Axis    & $E_{\textrm{exp}}$ & $H_{\textrm{exp}}$ & $W_{\textrm e}$  & Ref.\\
& & (GPa)&(GPa)&(\%)&\\\hline
\multirow{2}{*}{Zn{[}Au(CN)$_2${]}$_2$} & {[}100{]} & 16.8(3)   & 2.23(4)    & 62.7(3) &This \\ 
                                   & {[}001{]} & 22.1(4)   & 2.19(7)    & 43.6(6) &work\\ \hline
\multirow{4}{*}{CuPA1}            & {[}100{]} & 92.7     & 4.5   & 29.0 &\multirow{7}{*}{\citenum{Tan2009}}\\ 
                                  & {[}010{]} & 54.2   & 4.7    & 41.6 \\  
                                  & {[}001{]} & 49.8     & 4.5    & 49.8 \\
                                  & {[}011{]} & 57.3   & 4.2   & 45.8 \\                              
\multirow{3}{*}{CuPA2}            & {[}100{]} & 61.2     & 2.3   & 34.3 \\  
                                  & {[}010{]} & 34.5     & 2.5  & 45.4\\  
                                   & {[}001{]} & 55.2     & 2.3  & 33.1 \\ \hline
\multirow{3}{*}{CeOx}             & {[}100{]} & 51.8     & 4.1   & 40.5 &\multirow{3}{*}{\citenum{Tan2009a}} \\  
                                  & {[}010{]} & 43.0     & 3.9   & 46.4 \\  
                                  & {[}001{]} & 78.2     & 4.6   & 34.6 \\ \hline
\multirow{3}{*}{ZnPAA}  & {[}100{]}   & 53.3   & 3.4  & 48.0 &\multirow{3}{*}{\citenum{Kosa2010}} \\
                                                  & {[}010{]}  & 42.6      & 3.84   & 35.4   \\
                                                   & {[}001{]} & 49.6 & 3.0     & 40.4  \\ \hline
\multirow{2}{*}{ZIF-zni}          & {[}100{]} & 7.5    & 1.1    & 69.8(6) &\multirow{8}{*}{\citenum{Tan2010}}\\ 
                                   & {[}001{]} & 8.5    & 1.1    & 66.7(3)\\ 
ZIF-4                         & {[}111{]}  & 4.7   & 0.5  & 48.7(2)   \\                             
ZIF-7		        & {[}101{]}   & 6.2  & 0.65  & 50.5(9)  \\
ZIF-8                         & {[}001{]}  & 3.0  & 0.5    & 69.9(5)   \\ 
ZIF-9                        & {[}101{]}  & 5.9   & 0.7  & 59.4(14) \\ 
ZIF-20                       & {[}111{]}  & 3.9  & 0.25  & 28.9(6)  \\ 
ZIF-68                        & {[}001{]} & 3.7 & 0.35  & 52.8(8) \\ \hline                   
  \end{tabular*}
\end{table}

\begin{figure}
\centering
\includegraphics{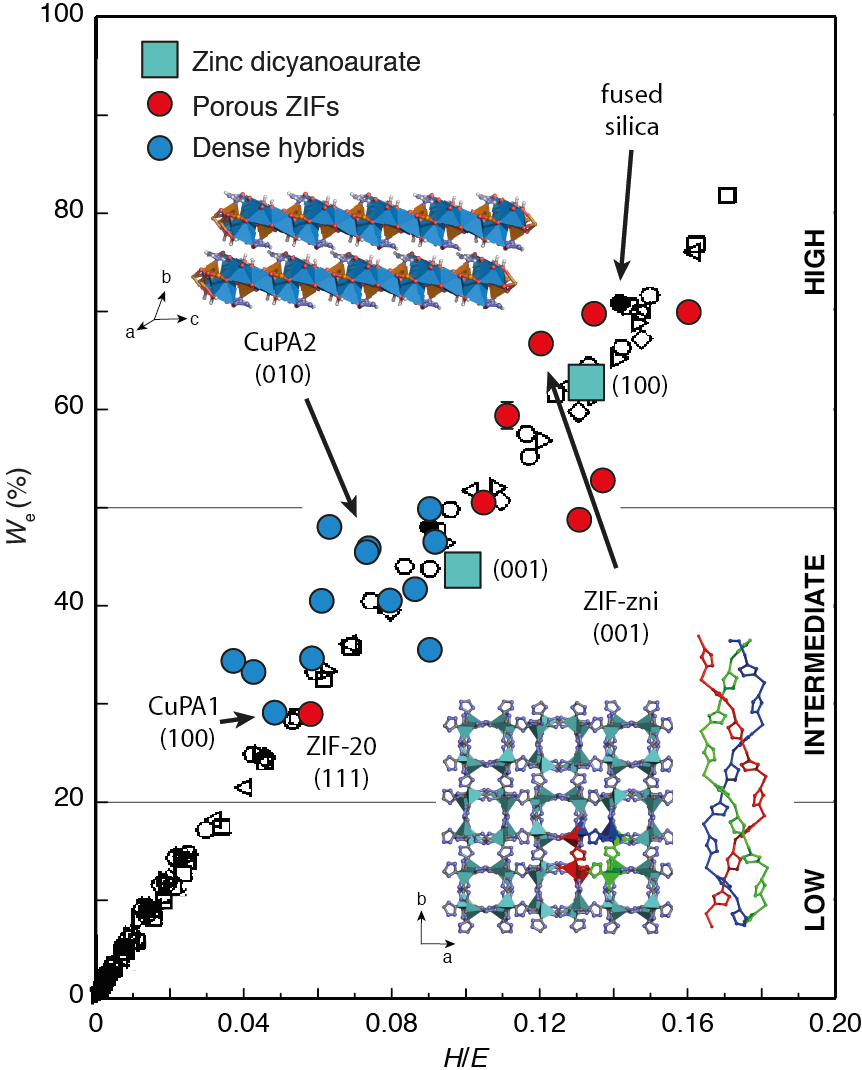}% Here is how to import EPS art
\caption{\label{fig3}Empirical relationship between elastic recovery $W_{\textrm e}$ and $H/E$. Data are given for Zn[Au(CN)$_2$]$_2$ (green symbols) and the variety of other porous (red symbols) and dense (blue symbols) framework materials for which data are given in Table~\ref{table1}. Also shown are the corresponding data for conventional materials as collated in Ref.~\citenum{Cheng1998} (black and white symbols). The data fall naturally into three approximate clusters: systems with low elastic recoveries (brittle ceramics), those with moderate values of $W_{\textrm e}$ (dense frameworks), and a small set of materials with high, polymer-like elastic recoveries. This last set includes many of the porous ZIFs. Zinc dicyanoaurate behaves as if a dense framework when stress is applied along the hexagonal $c$ axis, but as if an open MOF-type system when compressed along [100]. The insets show representations of the structures of ZIF-zni and CuPA2 as discussed in the text.}
\end{figure}

It has been shown previously that there is a linear empirical relationship between $W_{\textrm e}$ and the ratio $H/E$ of hardness to Young's modulus.\cite{Cheng1998,Bolshakov1998,Oliver2004} In Fig.~\ref{fig3} we show that this relationship appears to hold also for the data we list in Table~\ref{table2}, despite our approximation regarding indentation and Young's moduli. We also include in this figure for comparison the original data of Ref.~\citenum{Cheng1998} which makes clear just how exceptional the elastic recovery response of Zn[Au(CN)$_2$]$_2$ really is.

A corollary of the dependence of $W_{\textrm e}$ on $H/E$, rather than either hardness $H$ or compliance $1/E$ by themselves, is that compounds with remarkably different elastic moduli can nevertheless show the same degree of elastic recovery. For example, CuPA1 has an elastic modulus $E_{100}\,=\,93$\,GPa and ZIF-20 a modulus $E_{111}$\,=\,3.9\,GPa, and yet despite this order-of-magnitude difference their elastic recoveries are almost identical: $W_{\textrm e}=29$\% and 30\%, respectively.\cite{Tan2009} Hence it is meaningful to seek to characterise the elastic recovery of a compound irrespective of whether its intended application requires hardness and non-compliance or otherwise. 

 %\cite{Tan2010}\cite{Tan2009a}\cite{Tan2012}.

We suggest it is the supramolecular helix motif---instrumental in allowing NLC\cite{Baughman1998,Cairns2013,Cairns_2015}---that is also responsible for the remarkably high elastic recovery of Zn[Au(CN)$_2$]$_2$ along [100]. As discussed in the introduction, the helix is an object with the particular mechanical property of being able to accommodate bulk compression without substantial compression of the material from which it is constructed. Hence compression of Zn[Au(CN)$_2$]$_2$ along the [100] direction can in principle be accommodated without substantial changes in any bonding interactions within the solid. In this way, the strain induced during nanoindentation is less likely to result in bond breaking and fracture than might otherwise be the case. Instead the supramolecular springs effectively store the energy transferred during indentation (\emph{i.e.} work done), and release this energy reversibly upon removal of the indenter tip.

%Instead we find that the strain of indentation as well as providing macroscopic resistance to plastic deformation upon indentation by using the gold springs to store mechanical energy (work done). Instead of allowing the applied force to break chemical bonds (plastic deformation), the springs store this energy, which is then released upon removal of the indenter tip.

If this is indeed the case for Zn[Au(CN)$_2$]$_2$ , then it is natural to question whether similar supramolecular motifs may be responsible for strong elastic recoveries in other materials. The data in Table~\ref{table2} reveal that ZIF-zni shows an elastic recovery ($W_{\textrm e}=69$\%) that is comparable to that found in Zn[Au(CN)$_2$]$_2$. At face value, it is perhaps unexpected that this particular member of the ZIF family should show a particularly large value of $W_{\textrm e}$ since it is both the densest and the least compliant ZIF: its moduli are $E=7.5$--8.5\,GPa.\cite{Lehnert1980,Bennett2010} The material crystallises in the tetragonal space-group $I4_1cd$ and within the network are chains of Zn$^{2+}$ cations and imidazolate linkers that run along the $\mathbf{c}$-axis, forming a triple-helical structure as shown in the inset to Fig.~\ref{fig3}.\cite{Lehnert_1980} By analogy with the arguments presented above, we suggest that this triple helix would also provide a high resistance to permanent deformation. Indeed this triple-helix motif is also found in collagen---a crucial supporting components of cartilage, ligaments, tendons, bone and skin---which itself has a a high elastic recovery.\cite{Aifantis2011}

%and this can also be attributed to a helical motif within the structure. ZIFs, or zeolitic imidazolate frameworks, are a Zn- or Co- based family of MOFs that show tetrahedral coordination around the metal centre coordinated by various imidazolate linkers. This family of ZIFs were investigated via nanoindentation in Ref.\cite{Tan2010}. They showed that the elastic modulus in this family varied proportionally with the density and solvent accessible volume (SAV). Interest in ZIFs has largely stemmed from their potential in gas storage and separation, due to their high-porosity, and as sensors due to their low elastic moduli and large reponse to stimuli. 

%The high-T polymorph of ZIF-4, known as ZIF-zni, is t  Although its high density reduces its compliance, it shows the highest elastic recovery of all the compounds studied besides ZIF-8: the elastic recovery was found to be $W_e$\,=\,69\% and $W_e$\,=\,66\% respectively. 
 
Helices need not be the only supramolecular motif capable of facilitating large values or the elastic recovery. We suggest that weak inter-layer interactions found in layered framework materials may also facilitate improved elastic recovery because compression can be accommodated without significant changes in covalent bond lengths. This point is supported by the anisotropy of elastic recovery in CuPA2, a layered material: $W_{\textrm e}=34$\% and 33\% for the inter-layer directions ([100] and [001]), but $W_{\textrm e}=45$\% for the layer stacking direction [010] [see inset to Fig.~\ref{fig3}]. Similar behaviour is clearly evident in the nanoindentation study of layered dimethylsuccinates.\cite{Tan_2012b} So again weak supramolecular interactions appear capable of storing mechanical energy in these chemically very different systems.

\section{Concluding Remarks}

In this study we have used nanoindentation measurements and \emph{ab initio} calculations to explore the mechanical properties of the key NLC material zinc dicyanoaurate. Our main finding is that the material exhibits a remarkably strong elastic recovery when deformed perpendicular to its hexagonal axis. While the hardness of Zn[Au(CN)$_2$]$_2$ is largely independent of crystallographic orientation, there is significant anisotropy in both the elastic moduli and elastic recovery.  We suggest that the helical arrangement of aurophilic interactions within the framework confers macroscopic resistance to permanent, plastic deformation. Moreover, we infer that a related triple-helical motif may give rise to a similar effect in ZIF-zni. Our results suggest not only that the unusual compressibility of Zn[Au(CN)$_2$]$_2$ might indeed be exploited reversibly in practical applications, but---more generally---that helical structures are an attractive motif in the design of resilient mechanical structures.

\section*{Acknowledgements}
C.S.C., J.A.H. and A.L.G. gratefully acknowledge financial support from the European Research Council (Grant 279705), the Leverhulme Trust (Grant RPG-2015-292), and the Engineering and Physical Sciences Research Council (EPSRC) RCUK (Grant No. EP/G004528/2). J.-C.T. would also like to acknowledge the E.P.S.R.C. (Grant Nos. EP/K031503/1 and EP/N014960/1) for the provision of research funding.  M.R.R. thanks the E.P.S.R.C. for a D.T.A. postgraduate scholarship, the Science and Technology Facilities Council (S.T.F.C.) CMSD Award 13-05 for additional funding, and the Rutherford Appleton Laboratory for access to the SCARF cluster.


\balance


%%%REFERENCES%%%
\bibliography{References} %You need to replace "rsc" on this line with the name of your .bib file
\bibliographystyle{rsc} %the RSC's .bst file

\end{document}